# A survey of energies from pure metals to multi-principal element alloys

Ruitian Chen[1], Evelyn Li[1], Yu Zou[1,*]

[1]Department of Materials Science and Engineering, University of Toronto, 184 College St, Toronto, ON M5S 3E4, Canada

*Corresponding author. E-mail address: mse.zou@utoronto.ca (Y. Zou)

## Abstract

In materials science, a wide range of properties of materials are governed by various types of energies, including thermal, physicochemical, structural, and mechanical energies. In 2005, Dr. Frans Spaepen used crystalline face-centered-cubic (fcc) copper as an example to discuss a variety of phenomena that are associated with energies. Inspired by his pioneering work, we broaden our analysis to include a selection of representative pure metals with fcc, hexagonal close-packed (hcp), and body-centered cubic (bcc) structures. Additionally, we extend our comparison to energies between pure metals and equiatomic binary, ternary, and multi-principal element alloys (sometimes also known as high-entropy alloys). Through an extensive collection of data and calculations, we compile energy tables that provide a comprehensive view of how structure and alloying influence the energy profiles of these metals and alloys. We highlight the significant impact of constituent elements on the energies of alloys compared to pure metals and reveal a notable disparity in mechanical energies among materials in fcc-, hcp- and bcc-structured metals and alloys. Furthermore, we discuss the energy relationships, the implications for structural transformations and potential applications, providing insights into the broader context of these energy variations.

## 1. Introduction

A large majority of metal elements exist as crystalline solids at room temperature. Most metals and alloys crystallize in one of three common structures: face-centered cubic (fcc), hexagonal close-packed (hcp), and body-centered cubic (bcc). In the article by Dr. Spaepen [1], he compiled energy tables for fcc copper, exploring a diverse range of energy in materials science, including thermal, physicochemical, structural, and mechanical energies. This work offers a new approach to examining crystalline materials from an energy perspective. Dr. Spaepen used straightforward yet classic methods to calculate these energies and delved into the influence of energy levels on structural transformations.

Inspired by his article [1], we aim to broaden the scope of the energy analysis to include a wider array of materials, ranging from pure metals to equiatomic multi-principal element alloys, and from fcc and hcp to bcc structures. We collect all necessary parameters of these materials from existing literature (Tables S1-S3 in the Supplementary Materials), and the data is acquired by means of a combination of experimental measurements and theoretical calculations (details in the Supplementary Materials). Utilizing this comprehensive dataset, we calculate a series of energies, including energy related to temperature scales, thermal energy, magnetic energy, supersaturation, phase transition energy, vacancy and surface energy, dislocation, elastic strain, and external load energy. The computation methods largely adhere to Spaepen's approach [1]. We construct new energy tables for a much wider range of materials, providing an intuitive tool for showcasing and comparing energy levels across various materials and their properties, thereby facilitating a deeper understanding of the impact of energy on material's behavior.

## 2. Materials and energies

Table 1 presents the materials selected and the corresponding energies explored. In this study, the fcc materials consist of Period 4 transition elements such as Cr, Mn, Fe, Co, and Ni; the hcp materials are made up of Group 3 and 4 elements, including Sc, Ti, Zr, and Hf; the bcc materials comprise Group 5 and 6 elements, such as Nb, Mo, Ta, and W. There are four types of energy based on their properties: thermal energy, physicochemical energy (including magnetic energy and supersaturation), structural energy (including phase transition, vacancy, surface, and dislocation energy); and mechanical energy (including elastic strain and external load energy). Due to the lack of convincing data, quinary alloys of hcp structures are not included. Besides, our discussion will focus solely on the structural and mechanical energies for hcp and bcc materials.

Table 1. Materials and energy types: (a) Material types; (b) Energy types.

| (a) material types | fcc | hcp | bcc |
|---|---|---|---|
| pure metals | Ni | Ti | W |
| | | Zr | Mo |
| | | | |
| binary alloys | NiCo | TiZr | NbMo |
| | NiFe | | |
| | | | |
| ternary alloys | NiCoCr | TiZrHf | NbMoTa |
| | NiCoFe | | |
| | NiFeCr | | |
| | | | |
| quaternary alloys | NiCoFeCr | TiZrHfSc | NbMoTaW |
| | | | |
| quinary alloys | NiCoFeCrMn | \ | NbMoTaWV |

| (b) energy types | | fcc | hcp | bcc |
|---|---|---|---|---|
| Thermal | temperature scale | √ | | |
| | thermal energy | √ | | |
| Physicochemical | magnetic energy | √ | | |
| | supersaturation | √ | | |
| Structural | phase transition energy | √ | | |
| | vacancy energy | √ | | |
| | surface energy | √ | | √ |
| | dislocation energy | √ | √ | √ |
| Mechanical | elastic strain energy | √ | √ | √ |
| | external load energy | √ | √ | √ |

In this paper, we start by delving into fcc materials and discussing each type of energy in detail. Then, we will examine several energy aspects specific to hcp and bcc materials. Subsequently, we will compare materials of different structures, investigate the relationship between various energies, and examine the structural transformations induced by different energies.

## 3. Energies for fcc metals and alloys

In this section, we compare the energy profiles of fcc metals and alloys, including pure Ni, binary alloys NiCo and NiFe, ternary alloys NiCoCr, NiCoFe and NiFeCr, quaternary alloy NiCoFeCr, and quinary alloy NiCoFeCrMn (Figure 1).

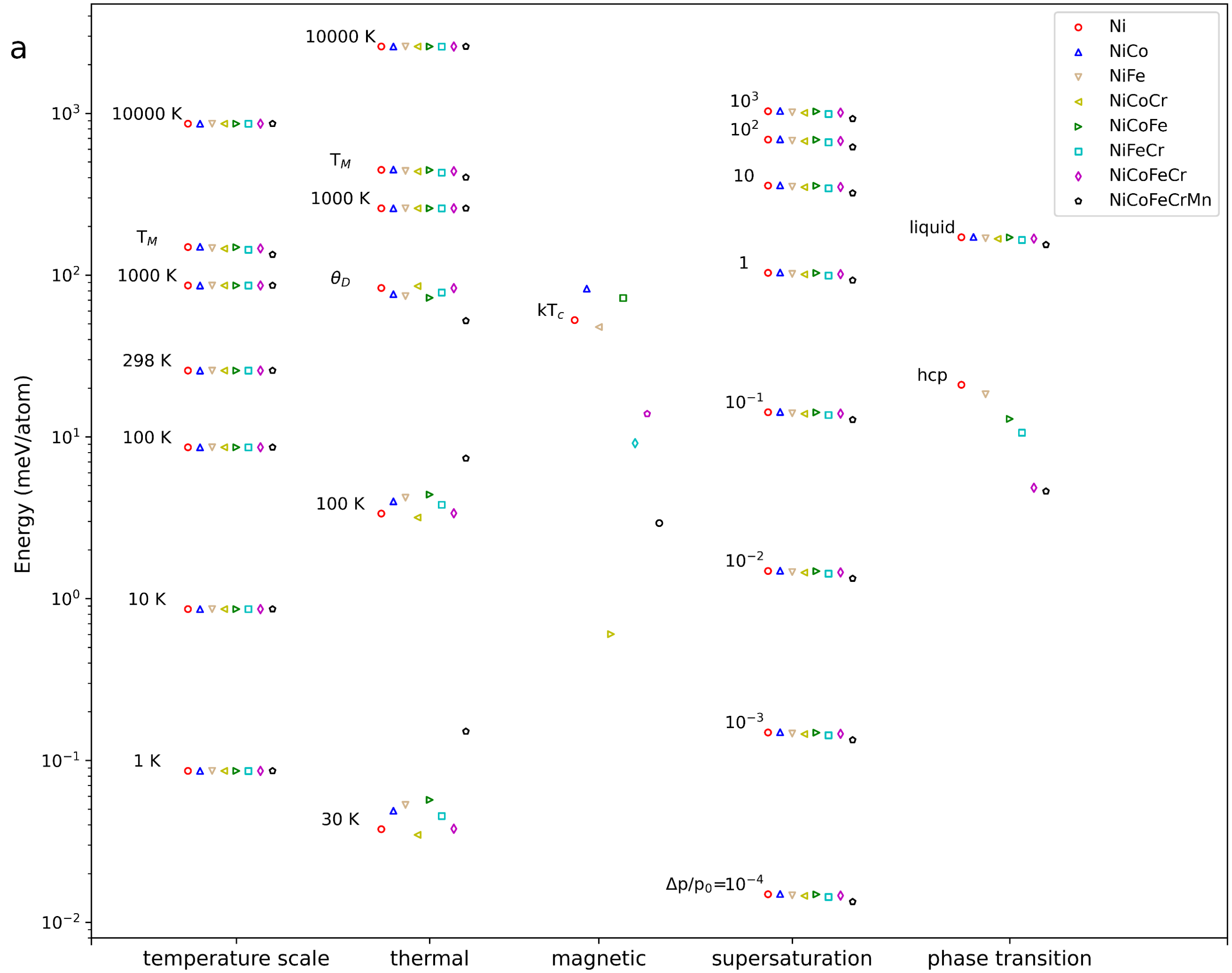
a
Energy (meV/atom)
temperature scale
thermal
magnetic
supersaturation
phase transition
Ni
NiCo
NiFe
NiCoCr
NiCoFe
NiFeCr
NiCoFeCr
NiCoFeCrMn
10000 K
$T_M$
1000 K
298 K
100 K
10 K
1 K
$\theta_D$
30 K
$kT_c$
$10^3$
$10^2$
10
1
$10^{-1}$
$10^{-2}$
$10^{-3}$
$\Delta p/p_0=10^{-4}$
liquid
hcp

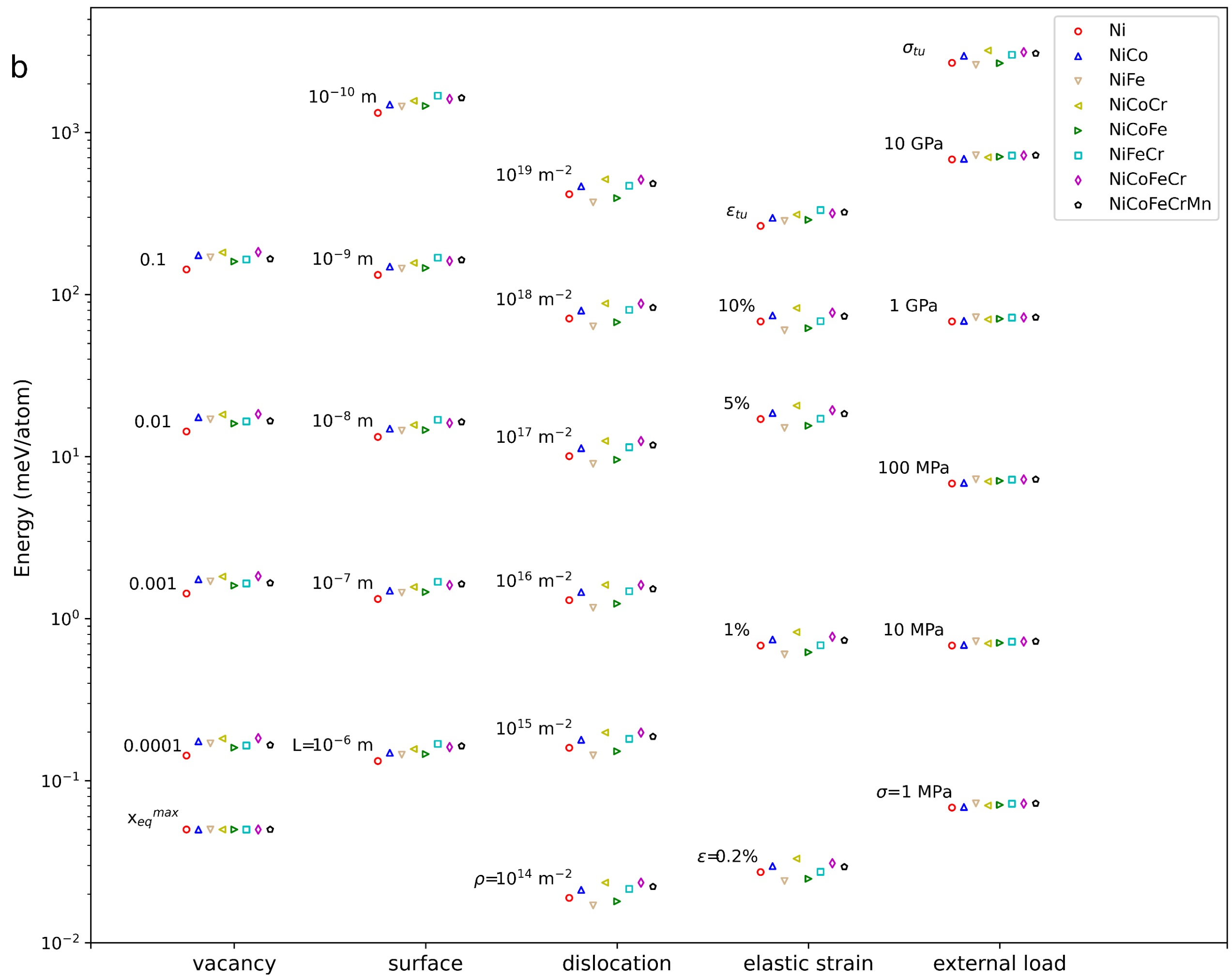

Figure 1. Energy tables for fcc metals and alloys including Ni, NiCo, NiFe, NiCoCr, NiCoFe, NiFeCr, NiCoFeCr, and NiCoFeCrMn. (a) Temperature scale and thermal energies at various temperatures, magnetic energy at the Curie point, chemical potentials at distinct supersaturation pressures, and phase transition energies for transformations from fcc to hcp structures and from solid to liquid states; (b) Vacancy energies at various vacancy fractions, surface energies at multiple crystal sizes, dislocation energies at varying dislocation densities, elastic strain energies at different strain levels, and external load energy under different loads.

### 3.1. Temperature scale and thermal energy

The equipartition theorem states that in the classical limit of statistical mechanics, the thermal energy of a system with $N$ independent degrees of freedom at thermodynamic equilibrium is given as follows [1,2]:

$$E_{\text{thermal}} = N\frac{k_{\text{B}}T}{2} \tag{1}$$

where $k_{\text{B}} = 1.381 \times 10^{-23}$ J/K is the Boltzmann constant and $T$ indicates the absolute temperature. In general, an atom in a condensed phase has three translational modes and three rotational modes, meaning the degree of freedom $N$ is six. Therefore, the thermal energy per atom is $3k_{\text{B}}T$, making $k_{\text{B}}T$ a natural thermal scale [1].

In Figure 1(a), the temperature scale shows the values of $k_{\text{B}}T$ for these fcc metals and alloys at different temperatures, ranging from 1 K to 10,000 K, including their melting points which are measured by the differential scanning calorimetry (DSC) methods [3-5]. The temperature $T$ roughly corresponds to an energy value of $0.1T$ meV/atom.

However, when $k_{\text{B}}T$ is smaller than the spacing between energy levels, the law of equipartition breaks down and quantum corrections must be considered. The Debye model, which estimates the phonon contribution to the heat capacity, predicts the low-temperature thermal energy of solids as [1,6]:

$$E_{\text{thermal}} = 9k_{\text{B}}T\left(\frac{T}{\theta_{\text{D}}}\right)^3\int_0^{\frac{\theta_{\text{D}}}{T}}\frac{x^3}{e^x-1}dx \tag{2}$$

where $\theta_{\text{D}}$ is the Debye temperature which is calculated from the crystal properties [7-12]. Debye temperature is the temperature at which the highest-frequency mode is excited and can be calculated briefly using [10]:

$$\theta_{\text{D}} = \frac{hv_{\text{m}}}{k_{\text{B}}} \tag{3}$$

where $h$ is the Plank constant and $v_{\text{m}}$ indicates the highest frequency of the atom.

If the temperature is above $\theta_{\text{D}}$, all the thermal energies are $3k_{\text{B}}T$, which obeys Dulong–Petit law [13]. At the melting point of these materials (around 1700 K), the energy level reaches above 400 meV/atom. Thermal energies become different if the temperature is lower than $\theta_{\text{D}}$. The $\theta_{\text{D}}$ of an alloy is expected to be lower than its constituents due to the local lattice strain caused by atomic size mismatch [14]. Nevertheless, Ni, Fe, and Co are right next to each other on the periodic table and have very similar atomic radii. Therefore, the $\theta_D$ values of their binary alloys are slightly smaller than those of pure Ni. The $\theta_D$ of the component elements also affects that of the alloy. Among ternary alloys, NiCoFe has the lowest $\theta_D$, because it contains no Cr which has a much higher $\theta_D$. Besides, the $\theta_D$ of NiCoFeCrMn quinary alloy is much lower than that of the others, which may be due to the low $\theta_D$ of Mn ($\theta_D$ of Ni, Co, Fe, Cr, Mn are 477 K, 460 K, 477 K, 606 K, 409 K, respectively [12]). Hence, the thermal energy of NiCoFeCrMn is much higher at the same temperature when accounting for quantum effects.

### 3.2. Magnetic energy

Due to the lack of data on the magnetic permeability for these alloys, the magnetic energy is estimated analogously to the thermal scale as follows [1]:

$$E_{\mathrm{magnetic}} = k_{\mathrm{B}} T_{\mathrm{c}} \tag{4}$$

where $T_c$ is the Curie point which is estimated within the mean field approximation [15]. $T_c$ is the temperature at which a ferroelectric material transitions from a low-temperature ferroelectric phase to a high-temperature paraelectric phase upon heating [16]. Thus, the thermal scale at $T_c$ reflects the energy level associated with magnetization. Additionally, $T_c$ is related to the sum of the magnetic interactions directly, and each element plays a vital role in it for alloys [15]. The Curie points of these materials vary considerably due to the distinct magnetic properties between the alloying elements. While Ni, Co, and Fe are all ferromagnetic, Cr is antiferromagnetic, and Mn is multi-magnetic [15]. If metals and alloys consist of only ferromagnetic elements, their Curie points are relatively high. However, the addition of Cr would decrease the Curie point significantly, because the existence of the antiferromagnetic Cr could make it easier to reduce the spontaneous magnetization.

Figure 1(a) clearly illustrates a significant disparity in the energy of magnetization, with a value of near 100 meV/atom for NiCo, while only around 1 meV/atom for NiCoCr. The stark contrasts in the energy of magnetization highlight the substantial variation in magnetic contributions among these materials.

### 3.3. Supersaturation

Supersaturation represents the deviation from the equilibrium [17]. When the vapor pressure ($p$) exceeds the equilibrium pressure ($p_0$) at the material's melting point, a change in chemical potential occurs as follows [1,17]:

$$\Delta\mu = k_B T_M \ln\left(1 + \frac{\Delta p}{p_0}\right) \quad (5)$$

where $\Delta p = p - p_0$ . For example, Figure S1 (Supplementary Materials) shows that supersaturation increases with overpotentials [17]. The values of supersaturation listed in Figure 1(a) can be easily achieved, indicating that evaporation and chemical vapor deposition (CVD) are also effective methods for producing high-energy metastable phases, such as amorphous materials [1]. This observation provides insight into the stability and behavior of these materials under supersaturation conditions. Furthermore, it is noteworthy that the melting points of Ni and its alloys are comparable [4,5]; hence, the variation in the chemical potential changes among them is minor, implying that these materials are likely to exhibit similar tendencies to exist in a supersaturated state at a given temperature.

### 3.4. Phase transition

For the energy of phase transition, the reference state here is fcc structure. Because the hcp structure is also one of the closest packed structures, it only shows a slight energy difference compared to the fcc structure. The hcp stack is of the "A-B-A-B" type, and the fcc stack has an "A-B-C-A-B-C" arrangement [18]. Thus, the hcp structure could be transformed from the fcc structure by introducing stacking faults between every other stack of close-packed planes. As the close-packed plane of fcc materials is (111), the phase transition energy from the fcc structure to hcp structure can be estimated as follows [1,19]:

$$E_{\text{HCP}} = \frac{1}{2}\gamma_{\text{SF}} A_{(111)} \quad (6)$$

where $\gamma_{\text{SF}}$ is the stacking fault energy at 0 K based on density functional theory (DFT) calculations [20-22], and $A_{(111)} = \sqrt{3}a^2/4$ is the area per atom in the (111) close-packed plane which can be derived from the lattice constant $a$ [3,7,23-25].

Figure 1(a) shows a decreasing trend of hcp energies as the number of components of materials increases, except for NiCo and NiCoCr, whose $\gamma_{\text{SF}}$ values are negative. According to Figure 1(a), although the hcp structure might be more stable than their fcc structure for NiCo and NiCoCr at 0 K, there are other factors influencing the structure stability, including vibration entropy and bond stiffness difference, especially at higher temperatures [22], so that at room temperature, NiCo and NiCoCr are still fcc structures. Since Co shows the hcp structure [26], the presence of Co in the material would lower the stacking fault energy. Among the ternary alloys in Figure 1(a), Cr shows a greater effect on decreasing $\gamma_{\text{SF}}$ than Fe. The addition of Mn, however, only has a minor impact on the quinary alloy. In general, the energy required for the phase transition from fcc structure to hcp structure is around 10 meV/atom which is not very high.

Compared to the change from one crystal structure to another, the transformation from solid to liquid state needs a considerably higher energy. Figure 1(a) displays the heat of melting ($\Delta H_{\mathrm{M}}$) of these materials. Here, $\Delta H_{\mathrm{M}}$ is derived from the entropy of melting ($\Delta S_{\mathrm{M}}$) [2]:

$$\Delta H_{\mathrm{M}} = T_{\mathrm{M}} \Delta S_{\mathrm{M}} \tag{7}$$

where $\Delta S_{\mathrm{M}} = 1.15 k_{\mathrm{B}}$, which includes $0.95 k_{\mathrm{B}}$ of increase in configurational entropy based on the polytetrahedral structure of the simple liquid, plus $0.2 k_{\mathrm{B}}$ increase in the vibrational entropy [1] according to a revised 'Richards rule' [1,27]. Therefore, the heat of melting is higher than 100 meV/atom which lines up with the temperature scale at the melting point.

### 3.5. Vacancy

Introducing vacancies in a lattice leads to an increase in the energy of the system; the energy can be estimated by the vacancy fractions and its formation energy as follows [1]:

$$E_{\mathrm{vacancy}} = xQ \tag{8}$$

where $Q$ is the average formation energy of vacancy based on DFT calculations [28-32] and $x$ the atom fraction of vacancies. The $Q$ depends on both composition and local environment, making it fluctuate in a range of about 0.5 eV. The local environment determines the energy required to remove an atom, and the chemical potential of the atom is determined by the composition [28]. There are only slight differences in the vacancy formation energies among these materials in Figure 1(a) (the coefficient of variation of $E_{\mathrm{vacancy}}$ for these materials is only about 7% at the same defect concentration), but the $Q$ values in alloys are generally larger than those in pure Ni, indicating that solid solutions reduce the vacancy concentration and make alloys more resistant to vacancy formation compared to pure metals [28].

The $E_{\mathrm{vacancy}}$ associated with the highest equilibrium vacancy fraction at the melting point is $x_{\mathrm{max}}^{\mathrm{eq}} Q = 0.05$ meV/atom [1], which is very small, as shown in Figure 1(b). Meanwhile, the upper limit of vacancy formation for the lattice stability is $x_{\mathrm{up}} = 0.1$ [1], corresponding to an energy level above 100 meV/atom which is comparable to the heat of melting.

### 3.6. Surface energy

The increase in energy due to the presence of interfaces and features of the material surface can be written as follows [1,2]:

$$E_{\text{surface}} = \frac{\gamma A \Omega}{V} = \frac{\gamma \Omega}{L} \tag{9}$$

where $A$ is the surface area, $V$ indicates the system volume, $\gamma$ represents the surface tension based on empirical equations or DFT calculations [33-37], and $\Omega$ denotes the atomic volume. For the fcc structure, $\Omega = a^3/4$ and $\boldsymbol{b} = a\,\langle\mathbf{110}\rangle/2$, where $a$ is the lattice constant and $\langle\mathbf{110}\rangle$ is the Miller index [38,39]. The surface energy is related to the surface area, and the surface-area-to-volume ratio $L = A/V$ expresses the linear dimension of the system, determining the surface energy. Figure 1(b) displays the surface energies of these materials across the range of $L$ spanning from $10^{-6}$ m to $10^{-9}$ m. Although no distinct trend emerges from our calculations, it is worth noting that, similar to vacancy energies, alloys generally exhibit slightly higher surface energies compared to pure Ni, possibly due to the solid solution strengthening for alloys [40].

Chemical short-range orders (CSROs), which refer to the non-random arrangement of atoms due to the complex interactions between the constituent elements, are an indispensable structural feature of medium or high-entropy alloys (M/HEAs) [41,42]. CSROs significantly affect the structural stability and properties of M/HEAs [41]. For instance, NiCoCr exhibits lower local surface energies as the CSRO degree increases, while the fluctuations in surface energy also decrease [43].

### 3.7. Dislocation

Dislocation density $\rho$ refers to the dislocation length per unit volume. The introduction of dislocation with a density of $\rho$ increases the energy of the lattice, and the relationship can be defined as [1,44]:

$$E_{\text{dislocation}} = \frac{|\boldsymbol{b}|^2 G}{4\pi} \ln\left(\frac{4}{|\boldsymbol{b}|\rho^{1/2}}\right) \Omega\rho \tag{10}$$

where $\boldsymbol{b}$ represents the Burgers vector, and $G$ the shear modulus obtained by various experimental tests [3,5,11,45].

Figure 1(b) displays dislocation energies of these materials with the $\rho$ from $10^{14}$ m$^{-2}$ to $10^{19}$ m$^{-2}$. Considering that the lattice constants of these materials are similar, the determinant of the dislocation energy is primarily the shear modulus. The shear modulus is influenced by the properties of the constituent elements (i.e., atomic size, bond strength), the concentration, and intrinsic interactions with the base element [46]. This non-linear relationship also transfers to the dislocation energy. For example, the dislocation energy does not exhibit an increasing trend when transitioning from quaternary NiCoFeCr to quinary NiCoFeCrMn. In addition, among ternary alloys, NiCoCr and NiFeCr exhibit about 7% and 16% higher dislocation energies than NiCoFe, respectively, at the same dislocation density. This disparity could be attributed to the fact that Cr has the most significant modulus mismatch when compared to the other elements, namely Ni, Co, and Fe, which possess more closely aligned shear moduli [45].

Theoretically, the maximum dislocation density can reach the areal atomic density $\Omega^{-2/3}$. Therefore, an upper limit of the dislocation density of these materials is estimated to be on the order of $10^{19}$ m$^{-2}$. As depicted in Figure 1(b), the maximum dislocation energy is approximately equivalent to the thermal energy at the melting point, surpassing 100 meV/atom. However, it is challenging to achieve a dislocation density as high as $10^{19}$ m$^{-2}$ in metals and alloys. For example, a transmission electron microscopy (TEM) image shows a high dislocation density of approximately $10^{16}$ m$^{-2}$ in a NiCoFeCrMn alloy (Figure S2 in Supplementary Materials) [47].

### 3.8. Elastic strain

The energy associated with a uniaxial strain ($\varepsilon$) is expressed as follows [1,48]:

$$E_{\text{strain}} = \frac{1}{2}\varepsilon^2 E\Omega \tag{11}$$

where $E$ is Young's modulus measured following the standard test method [3,5,11,45]. The maximum elastic strain ($\varepsilon_{\text{tu}}$) that corresponds to the ultimate tensile strength is related to the surface tension, as expressed by [1,49]:

$$\varepsilon_{\text{tu}} = \sqrt{\frac{\gamma}{E|\boldsymbol{b}|}} \tag{12}$$

At the upper limit of elastic strain, the energy can be expressed as follows:

$$E_{\text{strain,up}} = \frac{1}{2}\left(\sqrt{\frac{\gamma}{E|\boldsymbol{b}|}}\right)^2 E\Omega = \frac{\gamma}{2|\boldsymbol{b}|}\Omega \tag{13}$$

this energy value is independent of Young's modulus but is influenced by the surface tension (γ). Notably, the trend in the $E_{\text{strain}}$ closely resembles that of $E_{\text{surface}}$, underlining the interconnection between material deformation and surface properties.

When the elastic strain is below the maximum limit, Young's modulus becomes the dominant factor influencing the elastic strain energy. Young's modulus is also influenced by the properties of the individual elements constituting the material [45]. For example, the substantial mismatch in properties induced by Cr results in a higher elastic strain energy of NiCoCr than that of NiCoFe [45]. The comparative analysis of elastic strain energy provides insight into how materials respond to stress and strain.

### 3.9. External load

In the hydrostatic scenario, the energy associated with the work required to apply an external load to a system is related to the stress and the system volume, as expressed by [1,50]:

$$E_{\text{load}} = \sigma \Omega \tag{14}$$

where $\sigma$ is the external stress normal to a surface. Figure 1(b) shows the energies computed with a stress ranging from 1 MPa to the ultimate tensile strength ($\sigma_{\text{tu}}$), which can be calculated from the maximum elastic strain as follows:

$$\sigma_{\text{tu}} = E\varepsilon_{\text{tu}} = \sqrt{\frac{\gamma E}{|\boldsymbol{b}|}} \tag{15}$$

$\sigma_{\text{tu}}$ is affected by both the surface energy and Young's modulus. The values for $\sigma_{\text{tu}}$ of the materials shown in Figure 1(b) typically fall around 40 GPa. At the ultimate tensile strength, the external load energies could exceed $10^3$ meV/atom, offering a crucial insight into a material's mechanical robustness under significant stress conditions.

### 3.10. Summary for fcc metals and alloys

For the fcc metal and alloys discussed above, there is no uniform trend in their diverse properties regarding various energies, but the properties of a material often depend on its constituent elements and their interactions. For instance, the substantial disparity in the energy of magnetization is a direct result of the diverse magnetic properties exhibited by alloying elements. The presence of Co in a material can lower the phase transition energy from fcc structure to hcp structure, because of the inherent hcp structure of Co. On the one hand, the significant modulus mismatch between Cr and the other elements leads to higher dislocation energy and elastic strain energy of ternary

NiCoCr than those of NiCoFe. On the other hand, the higher vacancy energy and surface energy observed in alloys, compared to pure metal Ni, can primarily be attributed to the interactions between different elements within the alloy system.

## 4. Energies for hcp metals and alloys

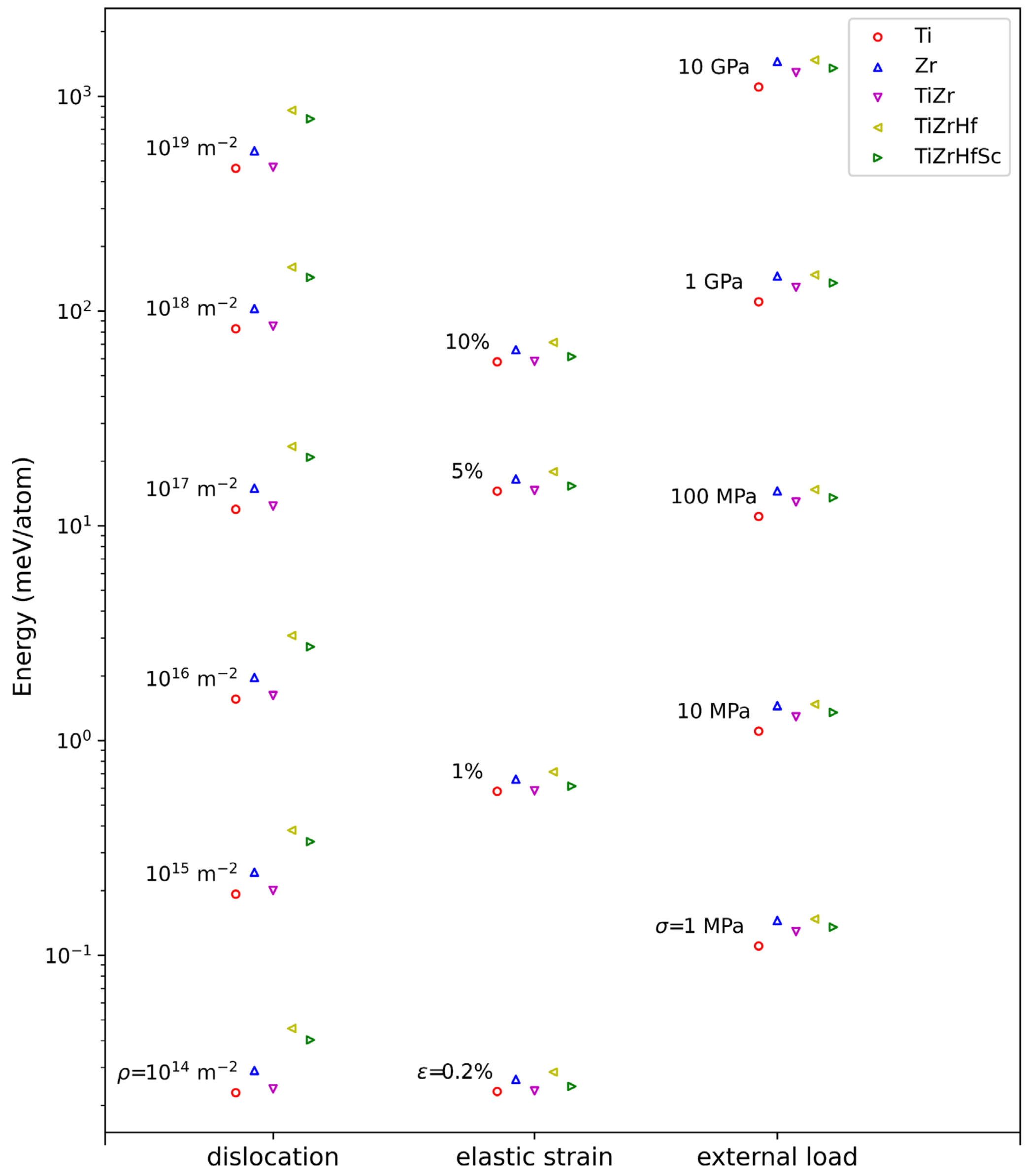

Figure 2. Energy tables for hcp metals and alloys: Ti, Zr, TiZr, TiZrHf, and TiZrHfSc. Dislocation energies at a wide range of dislocation densities, elastic strain energies at various strain levels, and external load energies under different loads.

The hcp lattice stands as another type of the closest packing structure. Unlike the fcc lattice, the hcp unit cell deviates from cubic symmetry and has two different lattice constants, denoted as $a$ and $c$. Although there are several pure metals in hcp structures, hcp multi-principal element alloys are rare. This may be because certain pure elements could exhibit an hcp structure at very low temperatures but would typically tend to transform to either bcc or fcc configurations at higher temperatures [51]. However, TiZr-based alloys are found to maintain the hcp structure [52,53]. In this section, we compare the dislocation energy, elastic strain energy, and external load energy of hcp metals and alloys, including pure Ti and Zr, binary alloys TiZr, ternary alloys TiZrHf, and quaternary alloy TiZrHfSc [52-59] (Figure 2). In the hcp structure, the atomic volume is determined as $\Omega = \sqrt{3}a^2c/4$, and Burger's vector is defined as $\boldsymbol{b} = a\langle \mathbf{11\bar{2}0} \rangle/3$ [60].

Illustrating the energy differences among hcp materials poses inherent complexities. Nevertheless, these materials share a common characteristic of having lower dislocation energies and elastic strain energies than fcc materials. This trend is partly attributed to the hexagonal arrangement which inherently possesses lower structural symmetry compared to cubic structures. Additionally, hcp materials generally exhibit larger atomic volumes, which, in turn, result in higher external load energies compared to fcc materials.

## 5. Energies for bcc metals and alloys

Different from the fcc and hcp structures, the bcc lattice is not a closest packing structure. In this section, we compare the energy profiles of bcc metals and alloys, including pure W and Mo, binary alloy NbMo, ternary alloy NbMoTa, quaternary alloy NbMoTaW, and quinary alloy NbMoTaWV. Due to limited available data, only surface energy, dislocation energy, elastic strain energy, and external load energy are obtained for this study [61-67] (Figure 3). The calculation formulas used for bcc materials are the same as those for fcc materials, except for the atomic volume ($\Omega$) that is determined as $a^3/2$ and Burger's vector that is expressed as $\boldsymbol{b} = a\langle \mathbf{111} \rangle/2$ for the bcc lattice.

The explanation for the energy differences observed in these materials is also analogous to that for the fcc materials above. For the surface energy, there is still no explicit trend. In terms of dislocation energy and elastic strain energy, the variations arise due to the modulus discrepancy among the constituent elements. For instance, the moduli of Nb and V are smaller than those of Mo and W [67]. Therefore, NbMo exhibits lower elastic strain energy than Mo. Similarly, NbMoTaWV demonstrates lower dislocation energy and elastic strain energy compared to NbMoTaW.

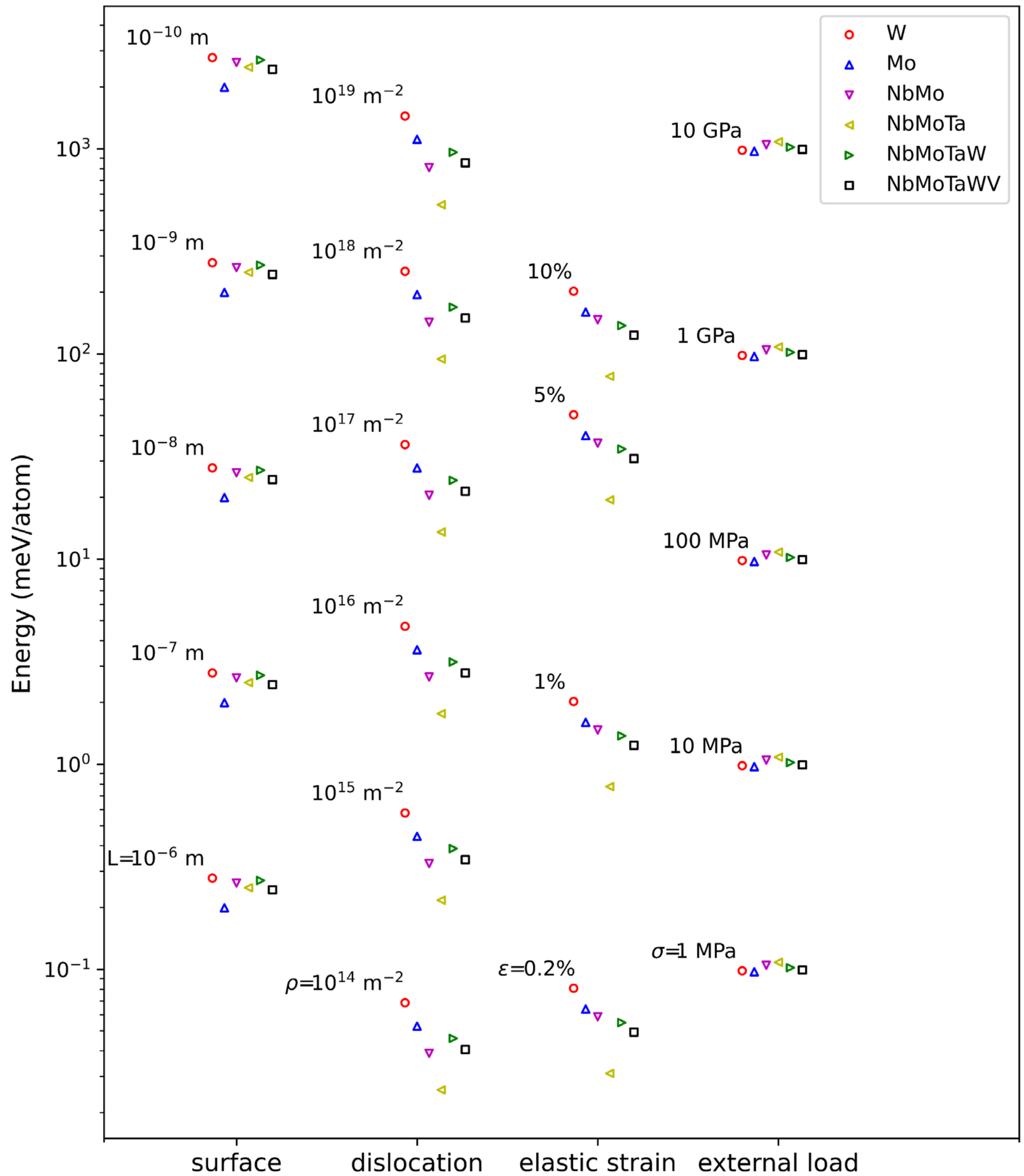

Figure 3. Energy tables for bcc metals and alloys: W, Mo, NbMo, NbMoTaW, and NbMoTaWV. Surface energies at different crystal sizes, dislocation energies at various dislocation densities, elastic strain energies at different strain levels, and external load energies under different loads.

## 6. Comparisons of materials in different structures

Figure 4 provides comparisons of dislocation energy, elastic strain energy, and external load energy across three pure metals and three quaternary alloys in three different crystal structures, respectively. Figure 4(a) compares Ni, Ti, and W, while Figure 4(b) shows the differences among NiCoFeCr, TiZrHfSc, and NbMoTaW.

The disparities in external load energy suggest that hcp materials possess larger atomic volumes than fcc and bcc materials. This observation underscores the critical role of atomic arrangement and symmetry on material properties and behaviors. The dislocation energy of bcc materials is approximately three times higher than that of fcc and hcp materials. Additionally, bcc materials exhibit elastic strain energy two times and ten times higher than that of fcc materials and hcp materials, respectively. The high dislocation energy and elastic strain energy of bcc materials could be because these bcc metals and alloys consist of refractory elements such as Nb, Mo, Ta, and W [68]. Refractory metals exhibit high interatomic bond strength and high strength; The strengths of refractory multi-principal element alloys surpass those of individual refractory metals [69,70]. Therefore, these refractory bcc materials, with higher Young’s modulus and shear moduli, show greater dislocation energy and elastic strain energy than fcc and hcp materials. In contrast, hcp materials exhibit the lowest elastic strain energy due to their small Young's modulus, as previously mentioned [52]. The correlation between structural properties and energies becomes apparent. These energy comparisons shed light on the distinctive mechanical behaviors and characteristics of these metals and alloys in different crystal structures.

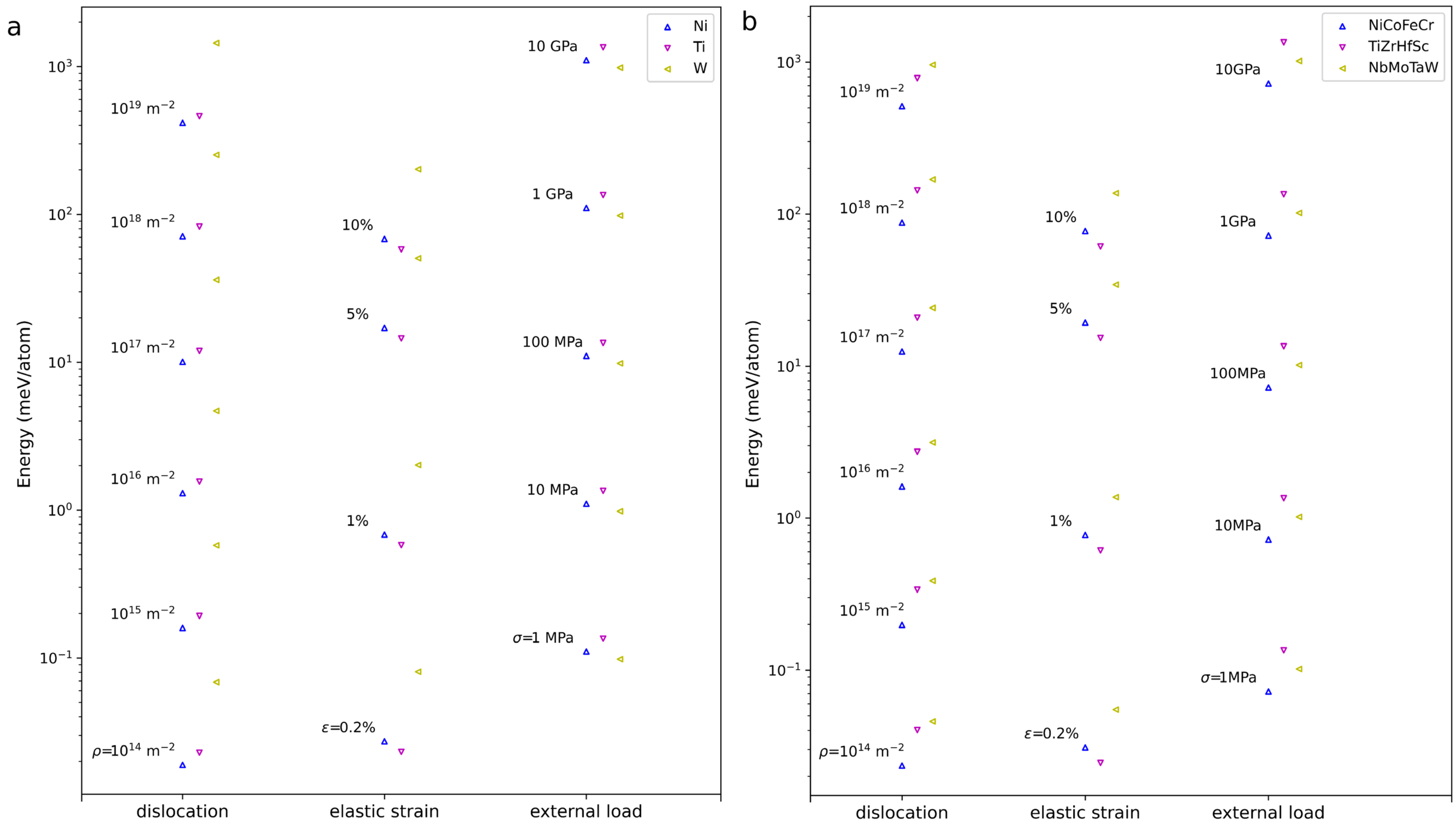

Figure 4. Energy comparisons including dislocation energy, elastic strain energy, and external load energy among materials in different structures. (a) Comparisons among pure metals: fcc Ni, hcp Ti, and bcc W; (b) Comparisons among quaternary alloys: fcc NiCoFeCr, hcp TiZrHfSc, and bcc NbMoTaW.

# 7. Discussions

## 7.1. Energy relationships

In solid materials, the conceptional framework of the Heckmann diagram provides an overview of the interrelationship between mechanical, thermal and electrical properties of materials. The Heckmann diagram highlights the coupled effects, such as piezoelectric interactions, where a small change in one property, such as mechanical stress, may produce a corresponding change in another, such as electrical polarization [71,72]. The interconnectedness underscores the importance of understanding and controlling these interactions to tailor material properties.

Modeled after the Heckmann diagram, we show the connections of different energies discussed in our study, as shown in Figure 5. The outer triangle denotes energies of configuration, while the inner triangle is constructed with respect to the energies of perturbation. The three vertices of the outer triangle are thermal, physicochemical, and mechanical energies, each of which exerts an influence on structural energies in the inner triangle. The parameters of the inner and the outer triangles have a direct correspondence respectively. For example, heat content is derived from temperature, while entropy determines the solid-liquid phase transition energy. Temperature and entropy can be related by heat capacity. Similarly, elasticity relates to stress and strain for mechanical energies. Different types of energies are coupled to each other, such as the thermostatic effects between temperature and stress, and the heat of deformation between strain and entropy. With regards to physicochemical and structural energies, magnetic fields or chemical potentials can cause structural changes to occur, such as vacancies and dislocations.

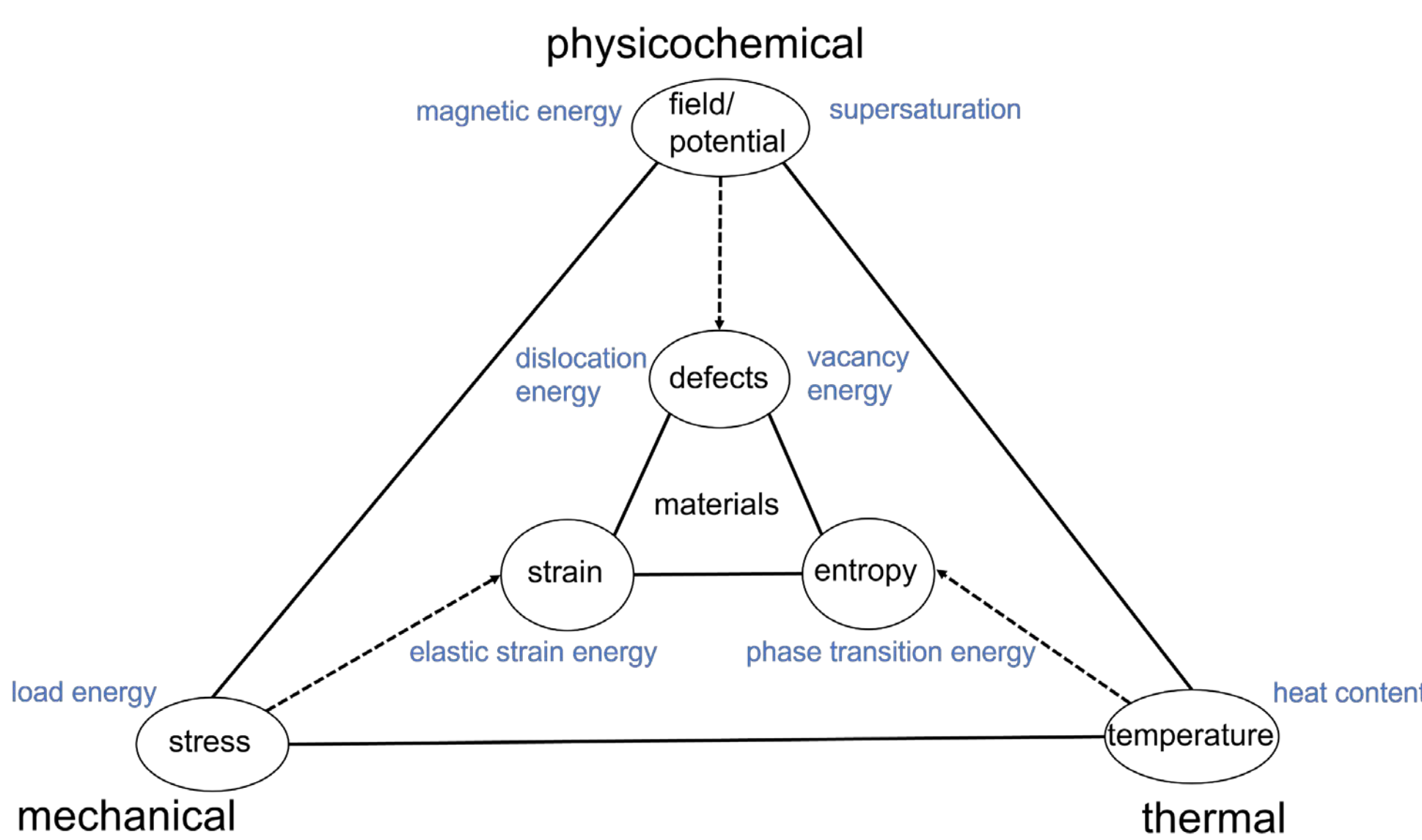

Figure 5. Energy diagram in materials science. The outer triangle denotes energies of configuration, while the inner triangle denotes energies of perturbation. Different types of energy interact and are interconnected.

By comparing the surface energy $E_{\text{surface}} = \gamma\Omega/L$ with the energy exerted by external forces $E_{\text{load}} = \sigma\Omega$, the ratio $\gamma/L$ can be regarded as the internal pressure within the material. The concept bears resemblance to the Laplace pressure commonly observed in fluid systems, where the pressure difference of a fluid is related to the surface tension and curvature of the interface [73]. There is a similar view in solid materials, where the internal pressure indicates the balance between the surface energy and the external forces, influencing the mechanical behavior of materials.

Comparisons between surface energy and supersaturation shown in Figure 1 reveal the relationship between the crystal size and the vapor pressure. Smaller-sized materials would exhibit higher supersaturation pressures. As the size of the material decreases, the surface-area-to-volume ratio increases, leading to an elevated surface energy. A higher surface energy needs a higher vapor pressure to achieve equilibrium, so the levels of supersaturation in nanomaterials are typically much higher than those in bulk materials, creating many unique properties of nanomaterials. Recently, the supersaturation-controlled surface structure strategy has emerged, involving the manipulation of growth unit supersaturation to control the surface structures of micro- and nanocrystallites. The surface energy of exposed facets is closely correlated with the supersaturation of growth blocks. Following this approach, micro- and nanocrystallites with diverse surface structures, particularly high-energy facets, have been successfully synthesized [17]. For example, the rapid generation of high levels of supersaturation with potent amorphous controlled precipitation nanoparticles offers new opportunities for improving bioavailability of poorly water-soluble drugs. The excellent wetting and rapid dissolution of the high-surface-area amorphous nanoparticles produce up to 90 times the equilibrium solubility, inhibiting solvent-mediated crystallization of the remaining solid drug in the presence of dissolution media [74].

### 7.2. Grain boundary energy

Grain boundary energy is the excess free energy associated with the presence of a grain boundary in polycrystalline materials. A grain boundary can be formed by creating two free surfaces and subsequently joining them. The grain boundary energy ($E_{\text{GB}}$) is less than twice the surface energy ($E_{\text{surface}}$) due to the binding energy ($B$) released when the two surfaces are combined and new bonds are formed [75]:

$$E_{\text{GB}} = 2E_{\text{surface}} - B \tag{16}$$

Grain boundaries can be categorized into low- and high-angle boundaries. Read-Shockley model successfully expresses the energy variation for low-angle boundaries as follows [76]:

$$E_{\text{GB}} = E_0\theta(A - \ln\theta) \tag{17}$$

where $\theta$ is the misorientation angle and $E_0$ and $A$ depend on the boundary plane orientation. A simple approach is to consider any high-angle boundary as a combination of pieces of two low-energy boundaries with different misorientations [77]. For example, grain boundary energies vary widely for fcc metals, and <111> twist and <100> twist boundaries are systematically low in energy. Typical values of average grain boundary energies vary from 0.32 J/m$^2$ for Al to 0.87 J/m$^2$ for Ni [78].

Grain boundary energies influence the thermal properties, mechanical properties, and grain sizes of materials. Grain boundary energy depends on temperature. For pure metals, it typically decreases linearly with increasing temperature due to the entropic term in the free energy. However, in alloys, the bulk solubility usually increases with temperature, causing solutes from intergranular regions to dissolve into the bulk and reducing boundary excess, and, consequently, the grain boundary energy increases [79]. Grain boundaries significantly affect the strength of materials by impeding dislocation as barriers or emitting dislocations as sources. The grain boundary energy, which captures the ability of grain boundaries to deform plastically, increases as the grain size decreases [80]. Additionally, grain boundary migration in nanograined metals contributes to the size dependence of strength under mechanical loading [81].

### 7.3. Structural transformations

In Figure 1(a), the energy level of the temperature scale at melting points and the heat of melting for the fcc metallic materials is around $10^2$ meV/atom, which can be considered a significant threshold for triggering structural transformations such as amorphization and nucleation of undercooled melt [82].

The vacancy energy can reach $10^2$ meV/atom when the vacancy fraction approaches its upper limit, as shown in Figure 1(b). However, such high vacancy concentrations are typically far from equilibrium and thus unlikely to occur except under ultimate conditions. Similarly, inducing structural transformations solely through dislocations would require unrealistically high dislocation densities exceeding $10^{18}$ m$^{-2}$ according to Figure 1(b). These extreme conditions are very rare in practical scenarios as discussed before, highlighting the challenge of achieving structural transformations through vacancy formation or dislocation mechanisms. For example, dislocation densities of $10^{12}$ m$^{-2}$ are commonly observed in metals and increase with plastic strain, but dislocation densities as high as $10^{16}$ m$^{-2}$ occur only in heavily deformed metals, and dislocation densities of $10^{18}$ m$^{-2}$ are almost impossible [83].

In terms of magnetic energy in Figure 1(a), the magnetic contribution of ferromagnetic materials can significantly influence structural transformations such as NiCo and NiCoFe, whose energies of magnetization are close to $10^2$ meV/atom. For instance, the NiCo/Ag multilayer films evolve from ferromagnetic to superparamagnetic with the increasing Ag layer thickness, accompanied by a transition from a layer system to a granular system consisting of NiCo giant particles [84]. The presence of ferromagnetic elements in these materials introduces magnetic interactions that can influence their behavior. In contrast, materials containing antiferromagnetic Cr with magnetic energy lower than 10 meV/atom are not eligible for magnetic-induced structural transformations.

Surface energy can be regarded as the thermodynamic driving force for sintering [85]. As demonstrated in Figure 1(b), when the size of clusters is reduced to the nanoscale level (For example, the surface energy for fcc gold nanoparticles could reach up to 7.7 J/$m^2$, which is around $10^3$ meV/atom [86]), the surface energy exceeds $10^2$ meV/atom, thereby making structural transformations feasible.

In Figure 1(b), it can be observed that the elastic strain energy is approximately $10^2$ meV/atom when the applied strain ($\varepsilon$) is 10%, which is below the maximum strain limit. Moreover, the energy input from external loads, measured in GPa is also capable of reaching levels up to $10^2$ meV/atom. Consequently, structural transformations in these materials can be triggered by either elastic strain or external forces. For instance, stress-assisted and strain-induced initiation of martensite phase transformations occurs in FeNiC alloys when applied stress reaches the order of 1 GPa [87].

Accumulating multiple types of energy can also potentially lead to structural transformations in materials. For example, while dislocation energy alone may not be sufficient to induce such transformations, the combination of dislocation energy, magnetic field, and external forces may have a synergistic effect on the material's behavior. The magnetic field may influence the alignment and the order of magnetic moments within the material, while the external forces can induce plastic deformation. The interplay between these energies may drive the material towards a different structural configuration. It is important to consider various factors that influence the specific outcome, including the material composition, the magnitude and orientation of the applied forces, and the interactions between different energy contributions. Nevertheless, it is important to acknowledge that the combined effects of different energies on structural transformations are still an area that requires further exploration.

### 7.4. Machine learning applications

Advanced computational methods, particularly machine learning (ML), offer new opportunities to obtain various energies in materials science and enhance our fundamental understanding of the correlations between them. For example, how these energies govern phase stability and transformations can be better understood by exploring connections among thermal, physicochemical, structural, and mechanical energies across fcc, hcp, and bcc structures in both pure metals and multi-principal element alloys. Recently, many models for predicting single-phase solid solution formation to identify molar volume, enthalpies, modulus, and melting temperature have been developed with the guidance of ML algorithm [88-90]. These studies exemplify an application of ML-informed feature selection that could directly enhance our own energy mapping across pure metals and complex alloys. By combining our comparative analysis of energies with ML approaches, the parameter space could be effectively expanded, improving phase stability predictions and providing a robust data-driven framework for understanding phase transformations across novel alloy compositions and crystalline structures. ML models have the potential to facilitate rapid screening of vast compositional spaces, identifying alloys with optimal phase stability and mechanical properties. The synergy between energy mapping and ML algorithms could accelerate the development of multi-principal element alloys.

## 8. Conclusions and outlook

In this article, we have compared various energies related to materials science for fcc, hcp, and bcc pure metals and multi-principal element alloys. This survey draws the following conclusions and outlook.

1) In our exploration of fcc metallic materials, we investigate their thermal, physicochemical, structural, and mechanical energies. The observed trends in energy profiles during the transition from pure metals to multi-principal element alloys are closely tied to the intrinsic properties of the constituent elements within the system. For example, the presence of antiferromagnetic Cr in the material significantly reduces its magnetic energy. Besides, the addition of Co leads to a decrease in the phase transition energy from an fcc structure to an hcp structure, owing to the inherent hcp structure of Co in its pure metal form. This observation indicates that when preparing alloy materials with specific desired properties, careful consideration must be given to the characteristics of their constituent elements. Selecting the appropriate elements based on their individual properties, such as atomic size, phase, and magnetism, is crucial to achieving the targeted mechanical and structural behaviors in alloys.

2) For hcp and bcc metals and alloys, we focus on their structural and mechanical energies. Similar to the fcc materials, the trends observed in hcp and bcc materials also suggest that the constituent elements play a pivotal role. For instance, the small modulus of one of the constituent elements in an alloy could significantly decrease its dislocation energy and elastic strain energy. This may influence the mechanical properties of an alloy, such as hardness and strength, demonstrating the importance of element selection in alloy design.

3) Regarding materials with different crystal structures, a distinct observation is that the energy gap is usually larger than the energy variations of materials within the same structure. For example, when comparing the elastic strain energies of fcc NiCoFeCr and fcc NiFeCr, the difference is only 5%. However, when comparing the elastic energies of fcc NiCoFeCr and bcc NbMoTaW, the difference is 60%. This trend demonstrates that while variations within the same crystal structure lead to slight energy differences, different crystal structures may have substantial impacts on the mechanical properties of metals and alloys.

4) There are close connections between various types of energies. For instance, elasticity links stress and strain in mechanical energies, while surface energy and supersaturation correlate with crystal sizes. Understanding these complex relationships not only enables informed decisions to optimize material performance and durability but also opens up practical applications. For example, the control of mechanical energies through elasticity can enhance material strength; the manipulation of surface energies and supersaturation may improve the thermal stability of nanocrystalline materials.

5) Grain boundary energy varies with temperature differently in pure metals and alloys, significantly influencing the mechanical properties and phase transformations of materials. Grain sizes, grain boundary angle, and grain boundary energy are essential factors in the development of polycrystalline metals and alloys for both structural and functional applications.

6) Structural transformations might be induced by several types of energies. For the fcc metals and alloys mentioned above, surface energy of nanoscale materials and high stress or elastic strain are feasible mechanisms for triggering structural transformations. From the perspective of energy, the interrelationships between different material properties can be revealed. By examining how energy is distributed and transformed within a material, this study offers a comprehensive framework to evaluate and predict the influence of variables on material properties.

7) With the rapid development of ML in materials science, the ML approach has been used extensively in the prediction of defect energy and phase transformation. Building upon this thorough survey of energies in a wide range of fcc, hcp and bcc metals and alloys, one promising direction of future work is the accelerated discovery of new materials and structures for targeted properties and performances. The extension of the dataset to include a wider range of materials and structures could contribute to the further enhancement of predictive accuracy in the ML models.

## DECLARATIONS

### Acknowledgments

The authors acknowledge Dr. Shuozhi Xu from the University of Oklahoma for some new data of bcc materials, including shear modulus of NbMo, surface energy of NbMoTa and NbMoTaWV, and Dr. Glenn Hibbard from the University of Toronto for constructive discussions.

### Authors' contributions

Conceived the idea and designed the project: Chen R, Li E, Zou Y
Performed data analysis and interpretation: Chen R, Li E
Supervised the project: Zou Y
Drafted the manuscript: Chen R
Revised and finalized the manuscript: Chen R, Li E, Zou Y

### Availability of data and materials

Supplementary Materials are available from the *Journal of Materials Informatics* or the authors.

### Financial support and sponsorship

The authors acknowledge the support from the Discovery Grants Program of the Natural Sciences and Engineering Research Council of Canada (NSERC) [RGPIN-2018-05731] and the Ontario Early Researcher Award (ER21-16-280).

### Conflicts of interest

All authors declared that there are no conflicts of interest.

### Ethical approval and consent to participate

Not applicable.

### Consent for publication

Not applicable.

### Copyright

## Supplementary Materials

# A survey of energies from pure metals to multi-principal element alloys

**Ruitian Chen, Evelyn Li, Yu Zou***

Department of Materials Science and Engineering, University of Toronto, 184 College St, Toronto, ON M5S 3E4, Canada

***Correspondence to:** Yu Zou, Department of Materials Science and Engineering, University of Toronto, 184 College St, Toronto, ON M5S 3E4, Canada. E-mail: mse.zou@utoronto.ca.

The parameters used to calculate energies are listed below.

Supplementary Table 1. Parameters of fcc materials

| | Ni | NiCo | NiFe | NiCoCr | NiCoFe | NiFeCr | NiCoFeCr | NiCoFeCrMn |
|---|---|---|---|---|---|---|---|---|
| lattice constant $a$ (Å) | 3.5238 [1] | 3.534 [2] | 3.595 [3] | 3.559 [4] | 3.569 [4] | 3.59 [5] | 3.592 [5] | 3.594 [5] |
| melting point $T_{\mathrm{M}}$ (K) | 1728 [1] | 1735 [6] | 1703 [6] | 1690 [6] | 1724 [6] | 1664 [6] | 1695 [6] | 1553 [7] |
| Debye temperature $\theta_{\mathrm{D}}$ (K) | 477 [8] | 437 [9] | 425 [10] | 490 [4] | 415 [4] | 448[#] [4] | 476 [11] | 299 [12] |
| Curie temperature $T_{\mathrm{c}}$ (K) | 611 [13] | 956 [13] | 554 [13] | 7 [13] | 838 [13] | 106 [13] | 161 [13] | 34 [13] |
| stacking fault energy $\gamma_{\mathrm{SF}}$ (J/m$^2$) | 0.125 [14] | -0.012 [15] | 0.105 [15] | -0.045 [15] | 0.075 [15] | 0.061 [16] | 0.0278 [16] | 0.0264 [16] |
| vacancy formation energy $Q$ (eV) | 1.47 [17] | 1.75 [17] | 1.70 [17] | 1.82 [18] | 1.60 [19] | 1.65 [20] | 1.83 [18] | 1.66 [21] |
| surface energy $\gamma$ (J/m$^2$) | 1.938 [22] | 2.161 [23] | 1.998 [24] | 2.231 [25] | 2.059 [25] | 2.343 [25] | 2.23 [26] | 2.26 [26] |
| shear modulus $G$ (GPa) | 76 [1] | 84 [7] | 62 [7] | 90 [27] | 68 [27] | 79 [27] | 86 [27] | 81 [12] |
| Young's modulus $E$ (GPa) | 200 [1] | 214 [27] | 166 [7] | 235 [27] | 175 [27] | 190 [27] | 214 [27] | 203 [12] |

(Note: $^{\#}\theta_{\mathrm{D}}$ = 448 K is for NiFe-20Cr)

Supplementary Table 2. Parameters of hcp materials

| | Ti | Zr | TiZr | TiZrHf | TiZrHfSc |
|---|---|---|---|---|---|
| lattice constant $a$ (Å) | 2.954 [28] | 3.232 [29] | 3.1141 [30] | 3.318 [31] | 3.147 [31] |
| lattice constant $c$ (Å) | 4.685 [28] | 5.147 [29] | 4.9221 [30] | 4.958 [31] | 5.05 [31] |
| shear modulus $G$ (GPa) | 41.4 [32] | 33.8 [33] | 33.5 [34] | 49.76 [31] | 52.97 [31] |
| Young's modulus $E$ (GPa) | 105 [35] | 91 [33] | 90.5 [30] | 96.9 [31] | 90.72 [31] |

Supplementary Table 3. Parameters of bcc materials

| | W | Mo | NbMo | NbMoTa | NbMoTaW | NbMoTaWV |
|---|---|---|---|---|---|---|
| lattice constant $a$ (Å) | 3.158 [36] | 3.1468 [36] | 3.224 [37] | 3.26 [38] | 3.195 [39] | 3.167 [39] |
| surface energy $\gamma$ (J/m$^2$) | 2.83 [40] | 2.05 [40] | 2.518 [41] | 2.31 | 2.661 [41] | 2.46 |
| shear modulus $G$ (GPa) | 161 [36] | 126 [36] | 82.57 | 51.87 [38] | 102 [39] | 94 [39] |
| Young's modulus $E$ (GPa) | 410 [42] | 329 [42] | 280.6 [37] | 143.69 [38] | 270 [39] | 249 [39] |

The various parameter values for the metals and alloy used to calculate these energies are sourced from a broad range of literatures. Differential scanning calorimetry (DSC) is used to quantify melting points for several Ni-based alloys, while shear and Young's modulus values are acquired using techniques such as resonant ultrasound spectroscopy and estimated from wave velocities. Lattice constants are determined using X-ray diffraction (XRD) and crystallographic databases, whereas Debye temperatures are derived from specific heat measurements and theoretical computations. Density functional theory (DFT) is extensively employed, especially when calculating the surface energies, vacancy formation energies, and other properties. Additionally, complex alloys are studied through methods like the special quasi-random structure (SQS) technique and virtual crystal approximation (VCA). Detailed methods are listed below:

## 1. Fcc materials

### 1.1. Lattice constant

- **Ni, NiCoCr, NiCoFe:** CRC Handbook and database [1,4].
- **NiCo**: Crystallographic data on electrodeposited phases and thin films [2].
- **NiFe**: Determined by XRD [3].
- **NiFeCr, NiCoFeCr, NiCoFeCrMn:** computed through supercells and replaced Ni by Co, Fe, Cr, and Mn simultaneously to build the crystal structure of the studied alloys [5].

### 1.2. Melting point

- **Ni:** CRC Handbook [1].
- **NiCo, NiFe, NiCoCr, NiCoFe, NiFeCr, NiCoFeCr, and NiCoFeCrMn:** Differential scanning calorimetry (DSC) measurements [6,7].

### 1.3. Debye temperature

- **Ni, NiCo:** Specific heat measurements [8,9].
- **NiFe:** Calculated from the shear and young's modulus measurements [10].
- **NiCoCr, NiCoFe, NiFeCr:** Calculated from atomic volume and the mean sound velocity [4].
- **NiCoFeCr, NiCoFeCrMn:** Database [11,12].

### 1.4. Curie temperature

- **Ni, NiCo, NiFe, NiCoCr, NiCoFe, NiFeCr, NiCoFeCr, and NiCoFeCrMn:** Calculation within the mean field approximation [13].

### 1.5. Stacking fault energy

- **Ni:** Anisotropic elasticity theory [14].
- **NiCo, NiFe, NiCoCr, NiCoFe, NiFeCr, NiCoFeCr, NiCoFeCrMn:** First-principles and DFT calculations [15,16].

### 1.6. Vacancy formation energy

- **Ni, NiCo, NiFe:** Computed through finite-size models while comparing to ab initio calculations to those obtained from available embedded atom method (EAM) potentials [17].
- **NiCoFe:** DFT calculations based on the Similar Atomic Environment (SAE) method [19].
- **NiFeCr:** Computed from atomistic calculations [20].
- **NiCoCr, NiCoFeCr:** DFT calculation on special quasi-random structures (SQSs) [18].
- **NiCoFeCrMn:** DFT supercell calculations [21].

### 1.7. Surface energy

- **Ni:** Computed through model which was conducted by a combination of Monte Carlo (MC), molecular dynamics (MD) and lattice statics (LS) types of simulations [22].
- **NiCo:** Taken from the weighted average of the measurement of surface tensions of molten Co–Ni alloys experimentally, also verified through theoretical calculations [23].
- **NiFe:** DFT calculations with dispersion correction [24].
- **NiCoCr, NiCoFe, NiFeCr:** DFT calculations verified through thermodynamic modeling and a bond-cutting model [25].
- **NiCoFeCr, NiCoFeCrMn:** DFT calculations performed using the Vienna ab initio Simulation Package (VASP) within the generalized gradient approximation (GGA) and with the Perdew-Burke-Ernzerhof (PBE) functional, also with the use of projector augmented wave (PAW) method [26].

### 1.8. Shear modulus

- **Ni:** CRC Handbook [1].
- **NiCo, NiFe:** Measured in ultrasonic techniques - resonant ultrasound spectroscopy [7].
- **NiCoCr, NiCoFe, NiFeCr, NiCoFeCr, NiCoFeCrMn:** Determined using torsional deformation of plates [12,27].

### 1.9. Young's modulus

- **Ni:** CRC Handbook [1].
- **NiFe:** Calculated from shear modulus and Poisson's ratio [7].
- **NiCo, NiCoCr, NiCoFe, NiFeCr, NiCoFeCr, NiCoFeCrMn:** The ASTM E1876-01 standard test method [12,27].

## 2. Hcp materials

### 2.1. Lattice constant

- **Ti:** Calculated by DFT based on the CASTEP [28].
- **Zr:** Special quasi-random structure (SQS) technique implemented in the Alloy Theoretic Automated Toolkit (ATAT) [29].
- **TiZr:** Determined by XRD [30].
- **TiZrHf, TiZrHfSc:** SQS - energy versus volume (E–V) curve fitting with the 3rd Birch–Murnaghan equation of state [31].

### 2.2. Shear modulus & Young's modulus

- **Ti:** Database [32,35].
- **Zr:** DFT on the basis of the frozen-core projected augmented wave (PAW) method of Blochl are performed within the Vienna ab initio simulation package (VASP), where the Perdew,

Burke, and Ernzerhof (PBE) form of the generalized gradient approximation (GGA) is employed [33].

- **TiZr: Shear modulus:** measuring the velocities of ultrasonic sound waves traveling in the samples. 40 **/Young's modulus:** Compressive mechanical properties were evaluated by testing machine (SANS CMT5504) [30].
- **TiZrHf, TiZrHfSc:** DFT, SQS for pristine structure [31].

## 3. Bcc materials

### 3.1. Lattice constant

- **W, Mo:** From experimentally determined atomic radius [36].
- **NbMo:** Determined by XRD [37].
- **NbMoTa:** DFT calculations [38].
- **NbMoTaW, NbMoTaWV:** Virtual crystal approximate (VCA) method [39].

### 3.2. Surface energy

- **W, Mo:** Used multiphase equilibration, computed from experimental values (experimental values are from geometric changes produced by vectorial interaction at intersecting interfaces) [40].
- **NbMo, NbMoTaW:** DFT calculations using the SQS supercells [41].
- **NbMoTa:** DFT calculations.

### 3.3. Shear modulus & Young's modulus

- **W, Mo:** Database [36,42].
- **NbMo: Shear modulus:** DFT. **/Young's modulus:** Oliver and Pharr method (indentation tests) [42].
- **NbMoTa:** DFT calculations [38].
- **NbMoTaW, NbMoTaWV:** Virtual crystal approximate (VCA) method [39].

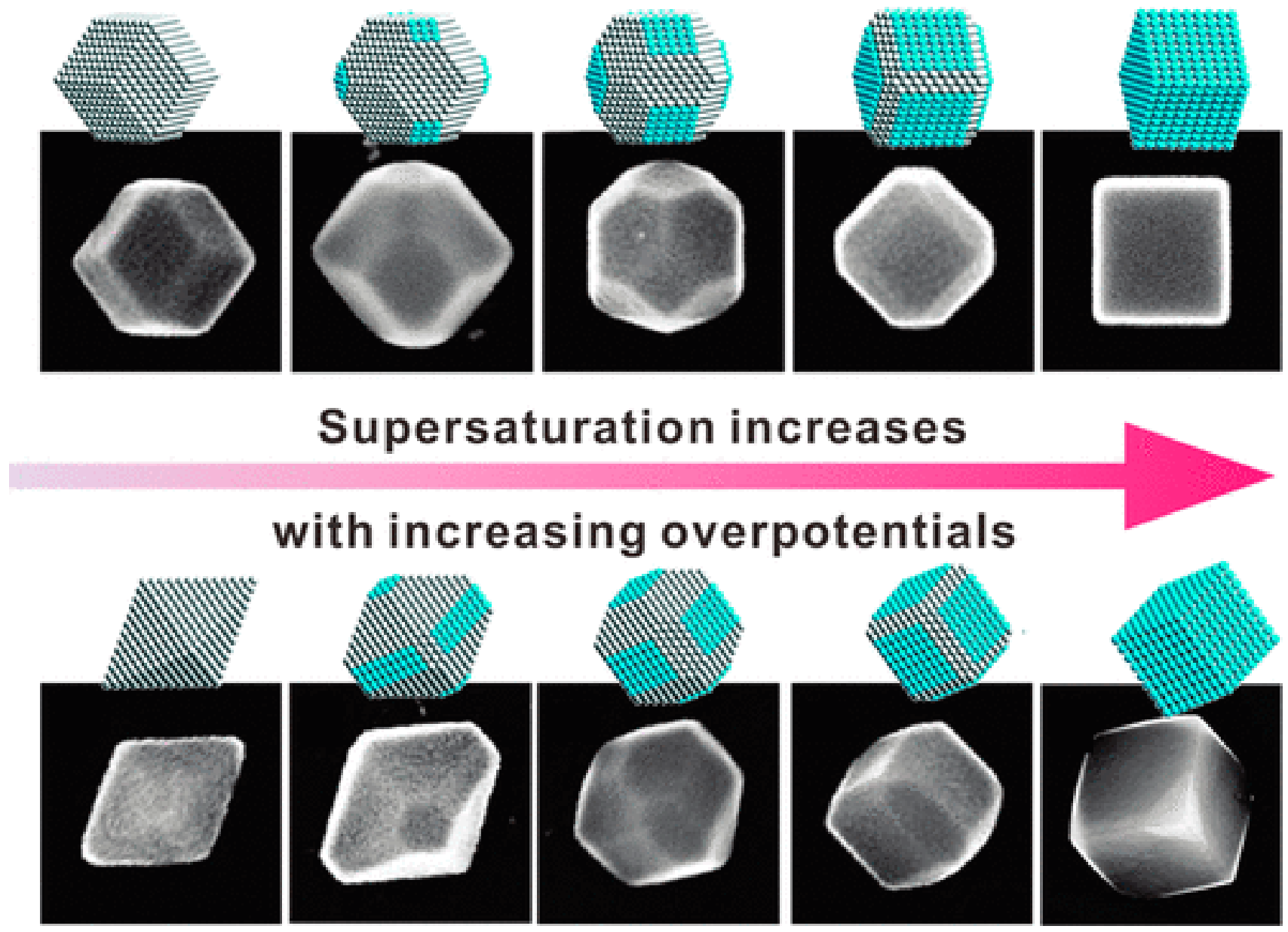

Supplementary Figure 1. Shape evolution of two series of Fe NCs from low-energy (110) facets to high-energy (100) facets with increasing the overpotential. Gray color, (110) facet; cyan color, (100) facet. Reprinted with permission from Ref. [43]. Copyright (2018) American Chemical Society.

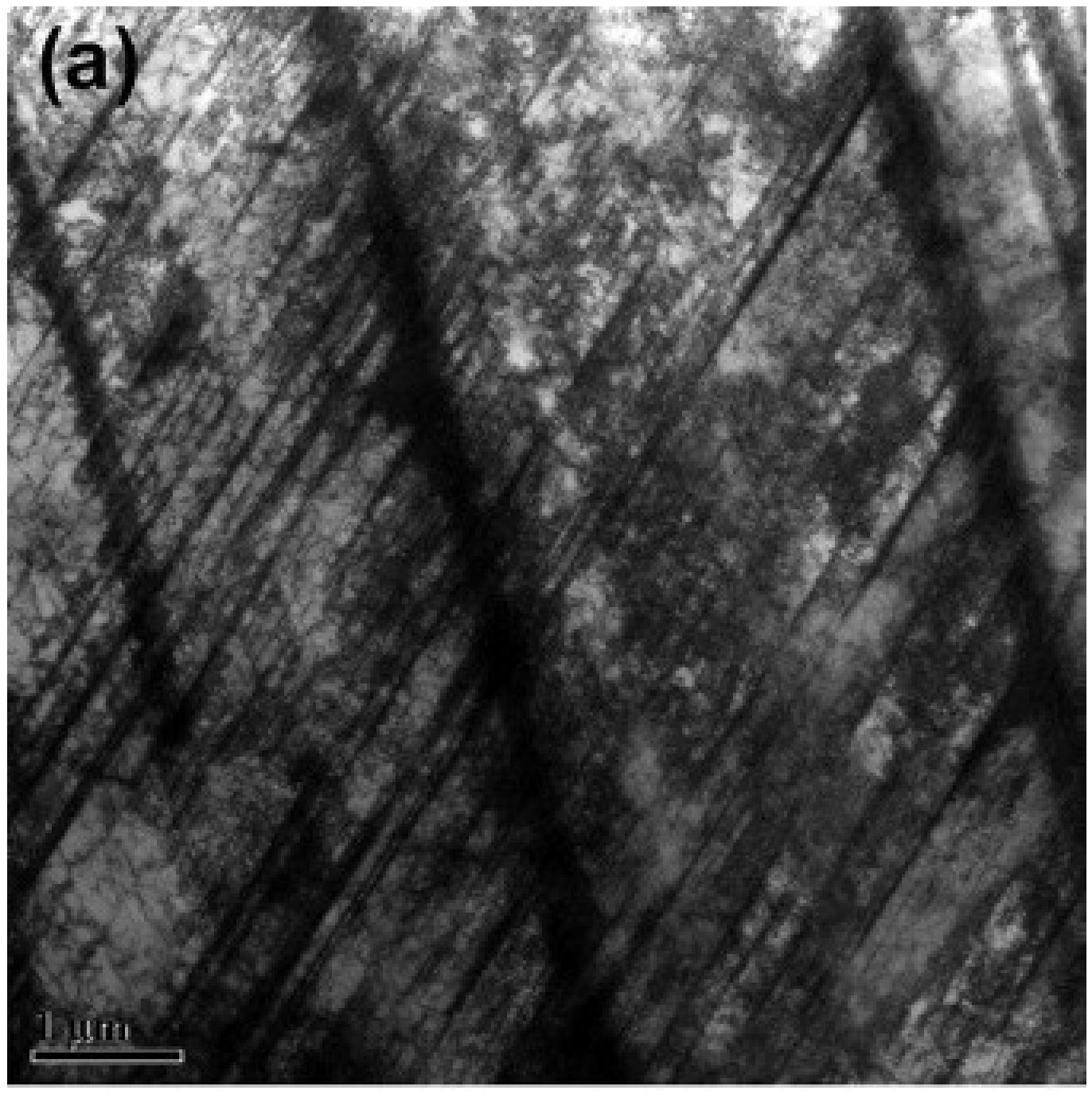

Supplementary Figure 2. Dark-field TEM image showing a high dislocation density of ~$10^{16}$ $m^{-2}$ in NiCoFeCrMn high-entropy alloy (Ref. [44]). Copyright (2020), with permission from Elsevier.